\documentclass[
aps,
amsmath,amssymb,
prb,
reprint,
superscriptaddress
]{revtex4-2}

\usepackage{graphicx}
\usepackage{dcolumn}
\usepackage{bbm}
\usepackage{bm}
\usepackage{hyperref}
\hypersetup{colorlinks=true,citecolor=blue,urlcolor=blue,linkcolor=blue}

\begin{document}


\title{Calibrating quantum hydrodynamic model for noble metals in nanoplasmonics}

\author{Qiang Zhou}
\altaffiliation{These authors contributed equally to this work.}
\affiliation{School of Physics and Wuhan National Laboratory for Optoelectronics, Huazhong University of Science and Technology, Luoyu Road 1037, Wuhan, 430074, China}
\affiliation{Institute for Quantum Science and Engineering, Huazhong University of Science and Technology, Luoyu Road 1037, Wuhan, 430074, China}

\author{Wancong Li}
\altaffiliation{These authors contributed equally to this work.}
\affiliation{School of Physics and Wuhan National Laboratory for Optoelectronics, Huazhong University of Science and Technology, Luoyu Road 1037, Wuhan, 430074, China}
\affiliation{Institute for Quantum Science and Engineering, Huazhong University of Science and Technology, Luoyu Road 1037, Wuhan, 430074, China}

\author{Zi He}
\affiliation{Department of Communication Engineering, Nanjing University of Science and Technology, Xiaolingwei 200, Nanjing, 210094, China}

\author{Pu Zhang}
\email[Corresponding author: ]{puzhang0702@hust.edu.cn}
\affiliation{School of Physics and Wuhan National Laboratory for Optoelectronics, Huazhong University of Science and Technology, Luoyu Road 1037, Wuhan, 430074, China}
\affiliation{Institute for Quantum Science and Engineering, Huazhong University of Science and Technology, Luoyu Road 1037, Wuhan, 430074, China}

\author{Xue-Wen Chen}
\email[Corresponding author: ]{xuewen\_chen@hust.edu.cn}
\affiliation{School of Physics and Wuhan National Laboratory for Optoelectronics, Huazhong University of Science and Technology, Luoyu Road 1037, Wuhan, 430074, China}
\affiliation{Institute for Quantum Science and Engineering, Huazhong University of Science and Technology, Luoyu Road 1037, Wuhan, 430074, China}

\begin{abstract}
Quantum hydrodynamic model (QHDM) has become a versatile and efficient tool for studying plasmonics at the nanoscopic length scale.
Yet its application to noble metals has not been sufficiently justified, in particular for situations where the metallic structures interface with dielectric material and electrons spill over the interfaces. In a recent work, we developed a refined QHDM, where the near-field effects and static polarization of metal ion lattice, and the electron affinity and static permittivity of the dielectric are incorporated. Here we perform a careful calibration of the model parameters for the refined QHDM. The model parameters are determined by benchmarking with (time-dependent) density functional theory calculations for special cases of simple metal. The predictive power of the refined QHDM with calibrated model parameters is faithfully demonstrated by the calculations of the optical responses from gold nanomatryoshkas of different sizes. The refined QHDM approach allows the quasinormal mode analysis for revealing the intrinsic optical properties of the nanoscopic metallic structures. We expect the well-calibrated refined QHDM would provide the nanoplasmonics community with a useful tool.
\vspace*{1.0cm}
\end{abstract}

\maketitle

\section{Introduction}
Plasmonic resonances of metallic nanostructures, i.e., collective oscillations of the conduction electrons, have the power of confining light down to 
\textcolor{black}{nanometer scale \cite{,Gramotnev2010Plasmonics,Baumberg2019ExtremeNanophotonics}.} 
Enabled by this unique capability nanoplasmonics spawn numerous applications in nano-optics and generates heated research interest \cite{,Thomas2003Surface,Schuller2010Plasmonics,Novotny2011Antennas,Polman2015Nanophoto,Chikkaraddy2016SingleMoleculeStrongCoupling,Liu2017StrongCoupling,Heiko2018StrongCoupling,Zhang2013RamanScattering,Benz2016Picocavities,Lee2019AtomicallyConfinedLight,Yang2020Photoluminescence,Qian2018IET,Parzefall2015IET,Hecht2015IET}.
When optical fields are concentrated towards the Thomas-Fermi screening length scale \cite{Ciraci2012PlasmonicEnhancement}, the quantum nature of conduction electrons becomes significant and needs to be treated properly. Theoretical methods at different levels of sophistication have been explored for proper description. Among them time-dependent density functional theory (TD-DFT) in principle provides exact description, where the conduction electrons in metal are characterized with the density $n(\mathbf{r},t)$ \cite{,Hohenberg1964DFT,Kohn1965DFT,Runge1984TDDFT} and current $\mathbf{J}(\mathbf{r},t)$ \cite{,Vignale2004MappingTDCDFT}.
The two fields are indirectly determined from the many-body Kohn-Sham (KS) equations \cite{,Kohn1965DFT,Runge1984TDDFT}, which are computationally very expensive to solve for plasmonic systems.
Alternatively, quantum hydrodynamic model (QHDM) emerges as a promising method that directly solves $\mathbf{J}(\mathbf{r},t)$ and $n(\mathbf{r},t)$ through a hydrodynamic equation \cite{,Toscano2015QHDM,Yan2015QHDM,Ciraci2016QHT,Ding2017QHDM,Ciraci2017CurrentDependent,Ciraci2021LaplacianLevel,Moldabekov2018QHD,Takeuchi2021Numerical}.
The equation retains the quantum and many-body effects with energy functionals \cite{,Zaremba1994Hydrodynamics,Ciraci2021LaplacianLevel} and viscoelastic terms \cite{,Ciraci2017CurrentDependent,Vignale1997TDDFT,Tokatly1999Hydrodynamic,Alvarez2020Hydrodynamic}.
In particular, electron spillover \cite{,Toscano2015QHDM} at metal surface is taken into account by a nontrivial ground-state density \cite{,Toscano2015QHDM,Yan2015QHDM,Ciraci2016QHT} as in TD-DFT.
In this regard, QHDM can account for the quantum effects with much cheaper computational cost, and thus outstands as a competitive choice for studying nanoplasmonics comparing with other semi-classical models \cite{,Luo2013LocalAnalogue,Christensen2017DParam,Yang2019DParam,PAD2020,Esteban2012QCM,Raza2011HWHDT,Mortensen2014GNOR}.

While QHDM has been well known to produce results in good agreement with TD-DFT calculations for simple metals such as sodium \cite{,Ciraci2017CurrentDependent}, the application to noble metals hasn't been systematically benchmarked.
A partial reason is that the treatment of metal in QHDM often relies on the traditional jellium model, where only the long-range Coulomb potential of the metal ion lattice is included.
The surrounding enviroment is also usually assumed vacuum.
In a recent work \cite{,Li2021BiasModulation} we filled the gap by developing a refined QHDM.
Besides the static permittivities of the surrounding dielectric and metal ion lattice \cite{,Khalid2020SecondHarmonic}, the refined QHDM introduces extra potentials to describe the near-field effects of metal ion lattice and electron affinity of the dielectric \cite{,Perdew1990StabilizedJellium,Fang2015QuantumSpillover}.
These factors prove crucial for determining the correct ground-state electron density distribution, especially in the spillover region, which in turn underlies the optical responses \cite{,Ciraci2016QHT}.
Nevertheless, different from first-principles theories, e.g. TD-DFT, there are free parameters in QHDM to be fixed according to other considerations.
In this work we carefully calibrate the model parameters with (TD-)DFT calculations and verify how the refined QHDM with the calibrated parameters perform for general cases. The remaining of this paper is structured as follows. The theoretical formulation of the refined QHDM is given in Sec.\,II.
Next in Secs.\,III and IV, we present the detailed procedures of the calibration. Benchmarks for the refined QHDM with the calibrated parameters are also demonstrated. Application of the refined QHDM to noble metal is exemplified in Sec.\,V.
Conclusions are finally drawn in Sec.\,VI.

\newpage
\section{Theoretical formulation}
\subsection{Formulation of quantum hydrodynamic model}
\textcolor{black}{
QHDM is essentially an orbital-free version of TD-DFT \cite{,Ciraci2021LaplacianLevel}, and the hydrodynamic equation of QHDM can be derived from the KS equations in TD-DFT \cite{,Ciraci2017CurrentDependent,Palade2018NonlocalOrbitalFree}.
In this work, we restrict the discussion to the stationary properties and linear optical responses.
So we drop the apparently high-order terms about the current $\mathbf{J}(\mathbf{r},t)$ and simplify hydrodynamic equation to
}
\begin{equation}\label{Eq:1:Hydro}
\frac{\partial\mathbf{J}}{\partial t}
= \frac{nq_\mathrm{e}^2}{m}(\mathbf{E}+ \!\frac{\nabla U_\mathrm{aff}}{q_e})
- \frac{nq_\mathrm{e}}{m}\nabla\frac{\delta G}{\delta n}
- \frac{q_\mathrm{e}}{m}\nabla\cdot(\eta_\sigma\sigma),
\end{equation}
where $q_\mathrm{e}$, $m$ and \textbf{E} are the electron charge, electron mass and electric field, respectively.
$G[n]=\int\!d\mathbf{r}\,g[n(\mathbf{r})]$ is the internal energy of the conduction electrons, with $g[n(\mathbf{r})]$ being the  energy density. 
\textcolor{black}{
Specifically, we have
\begin{align}
g[n] = t_\mathrm{TF}[n] 
+ t_\mathrm{W}[n] 
+ e_\mathrm{XC}[n],
\end{align}
where the first two terms constitute the Thomas-Fermi-von Weizs\"acker approximation of the kinetic energy density, and $e_\mathrm{XC}$ is Wigner's exchange-correlation (XC) energy density \cite{,Liebsch1997ElectronicExcitations,Yan2015QHDM,Ding2017QHDM}.} 
\textcolor{black}{The three energy densities read}
\begin{align}
t_\mathrm{TF}[n]
&= \frac{3\hbar^2}{10m}\,(3\pi)^{2/3}n^{5/3},
\\
t_\mathrm{W}[n]
&= \frac{\lambda_\mathrm{w}\hbar^2}{8m}\frac{\nabla n\cdot\nabla n}{n},
\\
e_\mathrm{XC}[n]
&= \left(\!\frac{0.035}{0.0625+7.8a_0n^{1/3}}-0.0588\!\right)
\frac{e^2n^{4/3}}{\varepsilon_0},
\end{align}
where $a_0=0.529$ \AA\ the Bohr radius, and $\varepsilon_0$ the vacuum permittivity.
Herein $\lambda_\mathrm{w}$ is the von Weizs\"acker parameter normally taken between $1/9$ and $1$.
The last term of Eq.\,\eqref{Eq:1:Hydro} characterizes nonlocal damping \cite{,Ciraci2017CurrentDependent} with the viscoelastic tensor
\begin{align}
\sigma_{\alpha\beta}
&= f_\mathrm{CV}\!\left(
\frac{\partial v_\alpha}{\partial r_\beta}
+\frac{\partial v_\beta}{\partial r_\alpha}
-\frac{2}{3}\delta_{\alpha\beta}\nabla\cdot\mathbf{v}\right)\!,
\\
f_\mathrm{CV}
&= \hbar n/\!\left(\!
60r_{s}^{-2/3}+80r_s^{-1}-40r_s^{-2/3}+62r_s^{-1/3}
\!\right)\!,
\end{align}
where $\mathbf{v}=-\mathbf{J}/(en)$ is the velocity field of conduction electrons, and $f_\mathrm{CV}$ is the Conti-Vignale interpolation function \cite{,Ciraci2017CurrentDependent,Conti1999Elasticity}.
$r_s=(4\pi n/3)^{-1/3}a_0^{-1}$ is the variable Wigner-Seitz radius.
The parameter $\eta_\sigma$ accompanying $\sigma$ is a free scaling factor which effectively controls the strength of nonlocal damping.
The affinity potential $U_\mathrm{aff}$ \cite{,Zhang2012BandBending} is additionally introduced in the refined QHDM recently developed by us \cite{,Li2021BiasModulation}.
It is piecewise constant in the metal and surrounding dielectric
\textcolor{black}{under flat-band approximation \cite{,Fang2015QuantumSpillover,Zhang2012BandBending}}.
While in metal $U_\mathrm{aff}=\langle\delta V\rangle$ denotes the near-field pseudopotential of the metal ion lattice \cite{,Perdew1990StabilizedJellium}, $U_\mathrm{aff}=U_{\mathrm{EA}}$ outside metal describes the electron affinity of the dielectric \cite{,Fang2015QuantumSpillover}.

In QHDM, the linear optical response is obtained following the standard perturbative treatment once we know the stationary properties.
The latter manifests the essential improvement of the refined QHDM and would be detailed later.
Assuming weak light excitation, the dynamic components then can be separated as \textcolor{black}{perturbative responses} from the stationary ones \cite{,Toscano2015QHDM,Yan2015QHDM,Ding2017QHDM,Ciraci2016QHT,Ciraci2017CurrentDependent}.
In particular, we have $n=n_0+n_1$, $\mathbf{J}=\mathbf{J}_1$ and $\mathbf{E}=\mathbf{E}_0+\mathbf{E}_1$.
The subscripts $0$ and $1$ denote static and dynamic components, respectively.
Substituting the perturbative expansions into Eq.\,\eqref{Eq:1:Hydro}, the linear response equation for $\mathbf{J}_1$ is found and written in frequency domain ($e^{-i\omega t}$ time convention) as
\begin{equation}\label{Eq:15:HEqV3}
(-i\omega+\gamma)\mathbf{J}
= -\frac{n_0q_\mathrm{e}}{m}\nabla\!\left(
\frac{\delta G}{\delta n}\right)_{\!1}
- \frac{q_\mathrm{e}}{m}\nabla\cdot(\eta_\sigma\sigma)
+ \frac{n_0q_\mathrm{e}^2}{m}\mathbf{E},
\end{equation}
where $(\cdots)_1$ means taking terms linear to $n_1$.
The phenomenological damping rate $\gamma$ is introduced to account for dissipation absent in TD-DFT.
Note that hereafter the subscript $1$ for $\mathbf{J}_1$ and $\mathbf{E}_1$ is dropped whenever no ambiguity arises.
Eq.\,\eqref{Eq:15:HEqV3} makes a closed theory by coupling with the continuity relation $q_\mathrm{e}\partial_tn_1+\nabla\cdot\mathbf{J}=0$ and the electric wave equation
\begin{align}\label{Eq:16:WaveEq}
\nabla\times\nabla\times\mathbf{E}
+ \frac{\omega^2}{c^2}\varepsilon_0\varepsilon_\mathrm{b}(\omega)\mathbf{E}
- i\omega\mu_0\mathbf{J}
= i\omega\mu_0\mathbf{J}_\mathrm{s},
\end{align}
where $c$, $\mu_0$ and $\mathbf{J}_\mathrm{s}$ respectively represent the speed of light, vacuum permeability and an external excitation.
For noble metals, besides the conduction electrons, the bound electrons also contribute to the optical responses.
The contribution is characterized here with \textcolor{black}{the permittivity function $\varepsilon_\mathrm{b}(\omega)$.}

\subsection{Refined ground state of conduction electrons}
The ground state shall be found by keeping the stationary terms of Eq.\,\eqref{Eq:1:Hydro}, \textcolor{black}{and satisfies}
\begin{align}\label{Eq:19:n0Origin}
\nabla\!\left(\!\frac{\delta G}{\delta n}\!\right)_{\!0}\!\!
- q_\mathrm{e}\mathbf{E}_0 - \nabla U_\mathrm{aff} 
= 0.
\end{align}
Here $(\cdots)_0$ means taking $n=n_0$.
The static electric field $\mathbf{E}_0=-\nabla\phi_0$ is solved self-consistently with \textcolor{black}{the Poisson equation:}
\begin{align}\label{Eq:18:phi0}
\nabla\cdot\varepsilon_0\varepsilon_\mathrm{r}\nabla\phi_0
= q_\mathrm{e}(n_+-n_0).
\end{align}
$en_+$ is the charge density of metal ion lattice, which is assumed uniform in jellium approximation \cite{,Perdew1990StabilizedJellium}.
Apart from $U_\mathrm{aff}$, the refined QHDM introduces the static permittivity $\varepsilon_\mathrm{r}$, which represents $\varepsilon_{\mathrm{ml}}$ of metal ion lattice in metal and $\varepsilon_\mathrm{d}$ in dielectric.
The static permittivities should be carefully assigned with appropriate (experimental) data.
Expressing $\mathbf{E}_0$ with $-\nabla\phi_0$, Eq.\,\eqref{Eq:19:n0Origin} is modified to
\begin{align}\label{Eq:19:n0}
\left(\!\frac{\delta G}{\delta n}\!\right)_{\!0}
+ q_\mathrm{e}\phi_0 - U_\mathrm{aff}
= \mu,
\end{align}
where $\mu$ is the chemical potential.
Practically, we further rewrite Eq.\,\eqref{Eq:19:n0} in terms of $f_0=(n_0/n_+)^{1/2}$.
Thereby the von Weizs\"acker kinetic energy term in $G$ produces a kinetic-energy-like operator $-\lambda_\mathrm{w}\hbar^2\nabla^2/(2m)$ acting on $f_0$.
Equation \eqref{Eq:19:n0} accordingly becomes a Schr\"odinger-like equation \textcolor{black}{for $f_0$:}
\begin{align}
&-\lambda_\mathrm{w}\frac{\hbar^2}{2m}\nabla^2f_0
+ V_\mathrm{eff}f_0
= \mu f_0, \label{Eq:f0}
\\
&\ V_\mathrm{eff}
= V_\mathrm{TF} + V_\mathrm{XC} + q_\mathrm{e}\phi_0 - U_\mathrm{aff},
\end{align}
The effective potential $V_\mathrm{eff}$ includes
$V_\mathrm{TF}=\delta T_\mathrm{TF}[n_0]/\delta n_0$
and
$V_\mathrm{XC}=\delta E_\mathrm{XC}[n_0]/\delta n_0$,
\textcolor{black}{
with
$T_\mathrm{TF}[n_0]=\int\!d\mathbf{r}\,t_\mathrm{TF}[n_0(\mathbf{r})]$
and $E_\mathrm{XC}[n_0]=\int\!d\mathbf{r}\,e_\mathrm{XC}[n_0(\mathbf{r})]$.}
The effective potential $V_\mathrm{eff}$ also gives the work function $W$ as \cite{,Yan2015QHDM}
\begin{align}\label{Eq:19:WorkQH}
W = V_\mathrm{eff}(\mathbf{r}\!\to\!\infty)
- \mu.
\end{align}
Eqs.\,\eqref{Eq:f0} and \eqref{Eq:18:phi0} together govern the ground-state electron density distribution.
When the plasmonic system is possibly electrostatically manipulated, e.g. charged or subject to a field $\mathbf{E}_\mathrm{c}$, the governing equations need to be adjusted correspondingly.
In the former case, the integral of the boundary flux of $\phi_0$ is constrained to fix the system's net charge.
In the latter scenario, $V_\mathrm{eff}$ should include an applied potential $\phi_\mathrm{ext}$ satisfying the homogeneous Poisson equation 
$\nabla\cdot\varepsilon_0\varepsilon_\mathrm{r}\nabla\phi_\mathrm{ext}=0$ 
with the boundary condition 
$-\nabla\phi_\mathrm{ext}|_{\mathbf{r}\!\to\!\infty}=\mathbf{E}_\mathrm{c}$.

\begin{figure}[b]
\centering
\includegraphics[width=0.48\textwidth]{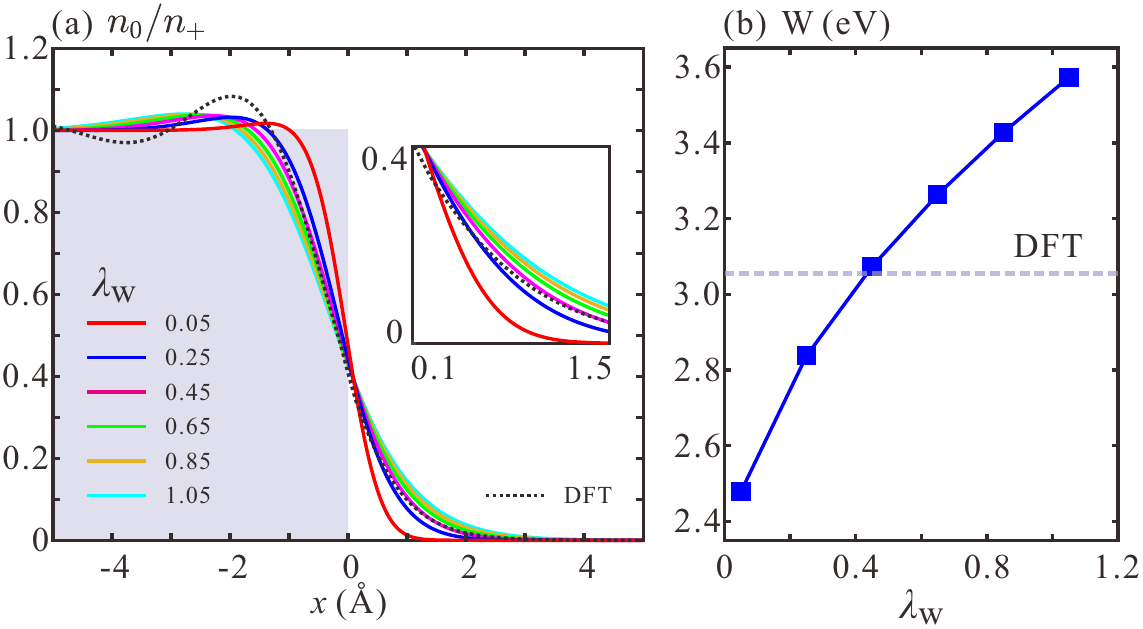}
\caption{
(a) Normalized stationary conduction-electron densities near the jellium edge ($x=0$) of a $5$ nm thick metallic slab with $r_\mathrm{s+}=4$ calculated with DFT (dashed) and QHDM using different $\lambda_\mathrm{w}$ (solid).
The shaded area denotes ionic charge distribution.
(b) Electron work functions predicted by DFT and QHDM.
The latter is a function of $\lambda_\mathrm{w}$.
}
\label{figV3StatPlt1}
\end{figure}
\begin{figure*}[!htb]
\centering
\includegraphics[width=0.95\textwidth]{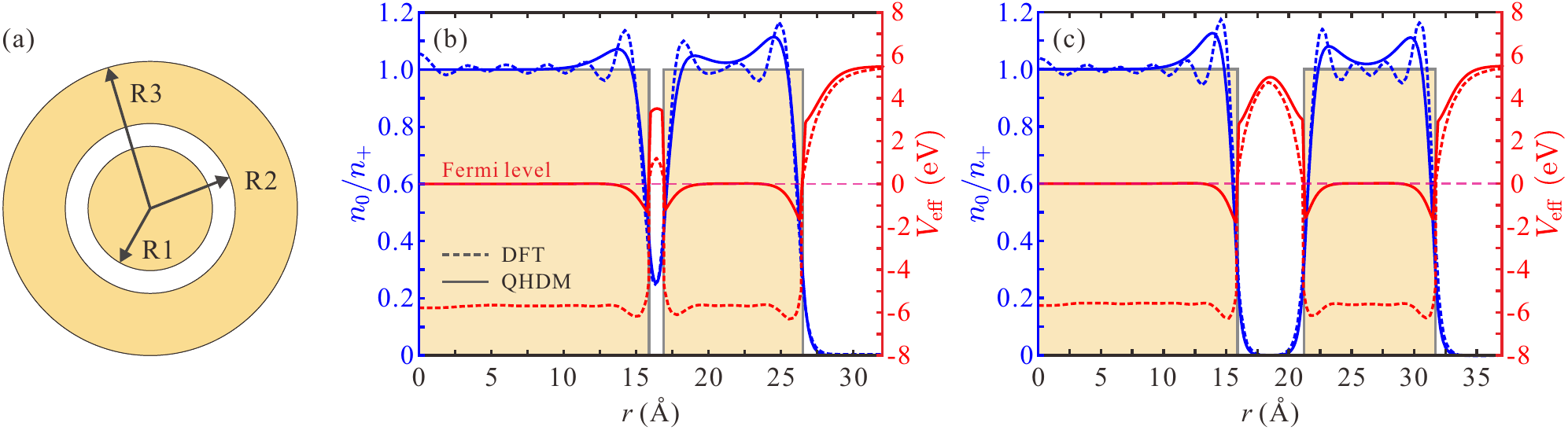}
\caption{
Ground states of gold nanomatryoshkas in vacuum calculated with DFT (dashed, data adapted from Fig.\,2 of Ref.\,\cite{,Kulkarni2013CoreShell}) and the refined QHDM (solid).
(a) Sketch of the geometry.
(b, c) Normalized stationary conduction-electron densities (blue) and effective one-electron potentials (red) for nanomatryoshkas of sizes 
$(R_1,R_2,R_3)
=(15.9,\allowbreak 16.9,\allowbreak 26.5)$ \AA\ 
and 
$(15.9,\allowbreak 21.2,\allowbreak 31.7)$ \AA\
respectively.
The shaded areas denote ionic charge distributions, and the Fermi levels are shifted to $0$ eV.
\label{figV3StatPlt2}}
\end{figure*}

\section{Calibration of ground-state QHDM}

In the ground-state equations \textcolor{black}{\eqref{Eq:f0} and \eqref{Eq:18:phi0}}, the von Weizs\"acker parameter $\lambda_\mathrm{w}$ appears as a model parameter to be fitted.
Since $\lambda_\mathrm{w}$ is a parameter with an approximate value ($1/9\!\sim\!1$),
\textcolor{black}{it is sensible to have different $\lambda_\mathrm{w}$ for the distinct physical situations of stationary state and optical responses \cite{,Yan2015QHDM}}.
In the ground state, $\lambda_\mathrm{w}$ is reflected in the spillover region of the electron density distribution, which plays a pivotal role in optical responses \cite{,Ciraci2016QHT}.
We thus aim to 
\textcolor{black}{calibrate $\lambda_\mathrm{w}$ by fitting $n_0$,} 
so that the $n_0$ profile in the spillover region can reproduce DFT results \cite{,Liebsch1997ElectronicExcitations}.
In view of the heavy computational load of DFT calculations, we choose to perform the calibration with a prototypical 1D metallic slab \cite{,Lang1970MetalSurfaceI,Lang1971MetalSurfaceII,Gao2006DFT1D,Neuhauser2012DFT1D,Fang2015QuantumSpillover}.

The metal is approximated with the simple jellium model in vacuum to keep the minimal factors for fitting $\lambda_\mathrm{w}$.
We assume the slab spans the interval $-L\leq x\leq 0$ with $L=5$ nm and the metal has $n_+\allowbreak =\allowbreak 3/[4\pi(r_{s+}a_0)^3]$ with the Wigner-Seitz radius $r_{s+}=4$ in atomic units.
The traditional jellium model with $r_{s+}=4$ is known to make correct predictions for sodium.
For comparison, we calculate the $n_0$ distribution of the slab with QHDM basing on Eqs. \eqref{Eq:f0} and \eqref{Eq:18:phi0}, and with DFT \cite{,Li2021BiasModulation}.
A series of values for $\lambda_\mathrm{w}$ is examined \textcolor{black}{in QHDM calculations}.
The $n_0$ distributions depicted in Fig.\,\ref{figV3StatPlt1}(a) show a good agreement between QHDM and DFT results 
\textcolor{black}{when $\lambda_\mathrm{w}\approx0.43$.}
Th fitting is confirmed by the work function plotted in Fig.\,\ref{figV3StatPlt1}(b) as a function of $\lambda_\mathrm{w}$.

Next we exemplify the ground state calculation with the refined QHDM for noble metal gold with $r_{s+}=3$.
In contrast with simple metal such as sodium, the bound electrons of gold have to be taken into account.
The polarization of gold ion lattice gives rise to 
\textcolor{black}{the static permittivity $\varepsilon_{\mathrm{ml}}=8.0$ \cite{,Kulkarni2013CoreShell}.}
The lattice's near-field pseudopotential $\langle\delta V\rangle=4.7$ eV is indirectly determined by requiring the resulting work function equal the 
\textcolor{black}{experimental data $5.4$ eV \cite{,CRC}.}
We concretely study the gold nanomatryoshka illustrated in Fig.\,\ref{figV3StatPlt2}(a).
The geometry is specified by the triplet of radii $(R_1,R_2,R_3)$.
The refined QHDM is employed to solve for the ground-state electron densities for two gold nanomatryoshkas of different sizes.
The resulting distributions (solid) ,of stationary electron density and effective potential are displayed in Fig.\,\ref{figV3StatPlt2}(b,\,c) in parallel with the DFT results (dashed) adapted from Ref.\,\cite{,Kulkarni2013CoreShell}.
For both nanomatryoshka structures, excellent agreement in the $n_0$ profiles is observed near metal surface and in the gap.
Notably, our QHDM exactly reproduces the electron tunneling across the $1$ \AA\ wide vacuum gap in Fig.\,\ref{figV3StatPlt2}(b).
In this case there's discrepancy in the potential barrier comparing with the DFT result.
Nevertheless, we emphasize that it's the $n_0$ distribution, rather than the effective potential, that enters the calculation of the optical responses as input.
In addition, the refined QHDM has been demonstrated 
\textcolor{black}{in our recent work \cite{,Li2021BiasModulation}}
to accurately produce $n_0$ distribution for various metals, e.g. gold, silver and aluminum, and in the presence of external static bias.
Hence the predictive power of the refined QHDM on the ground-state electron density is clearly appreciable when the near-field pseudopotential and static permittivity of metal ion lattice are properly treated.

\begin{figure}[b]
\centering
\includegraphics[width=0.48\textwidth]{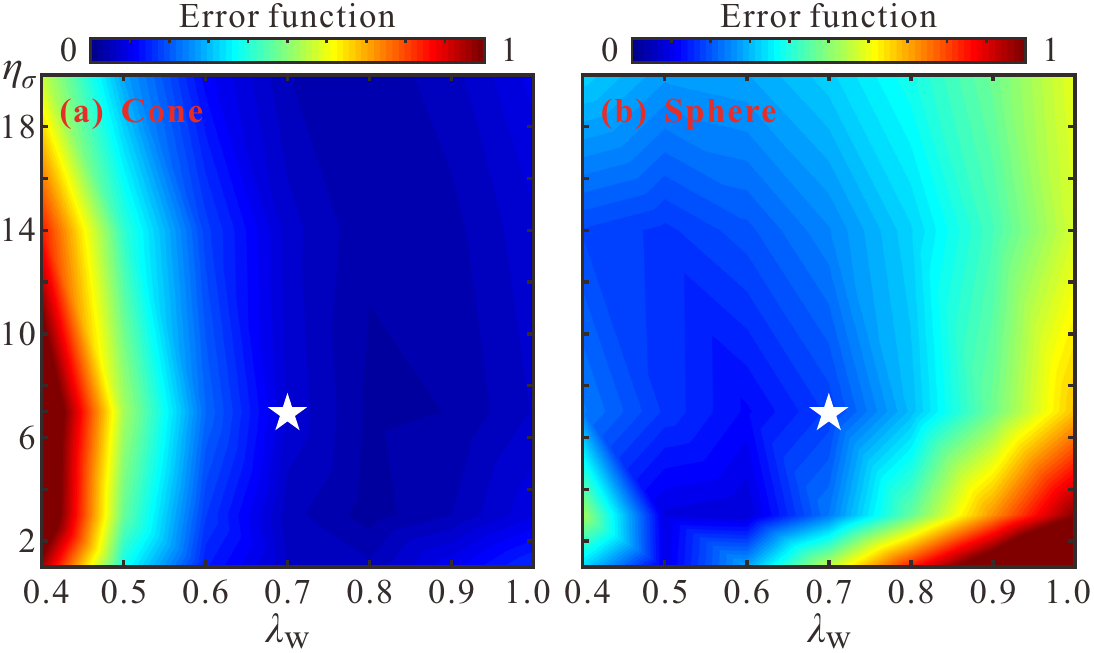}
\caption{
Error functions for assessing the predictive powers of the refined QHDM on linear optical response for (a) a sodium nanocone and (b) a sodium nanosphere subject to electrostatic manipulation.
Lower error indicates that the predicted frequencies and relative amplitudes of the major plasmon resonance are closer to the TD-DFT predictions in Fig.\,\ref{figV3SpecPlt2}(a, c); see Appendix C.
The white stars mark the optimal values of $\lambda_\mathrm{w}$ and $\eta_\sigma$.
\label{figV3SpecPlt1}}
\end{figure}

\begin{figure*}[htb]
\centering
\includegraphics[width=0.95\textwidth]{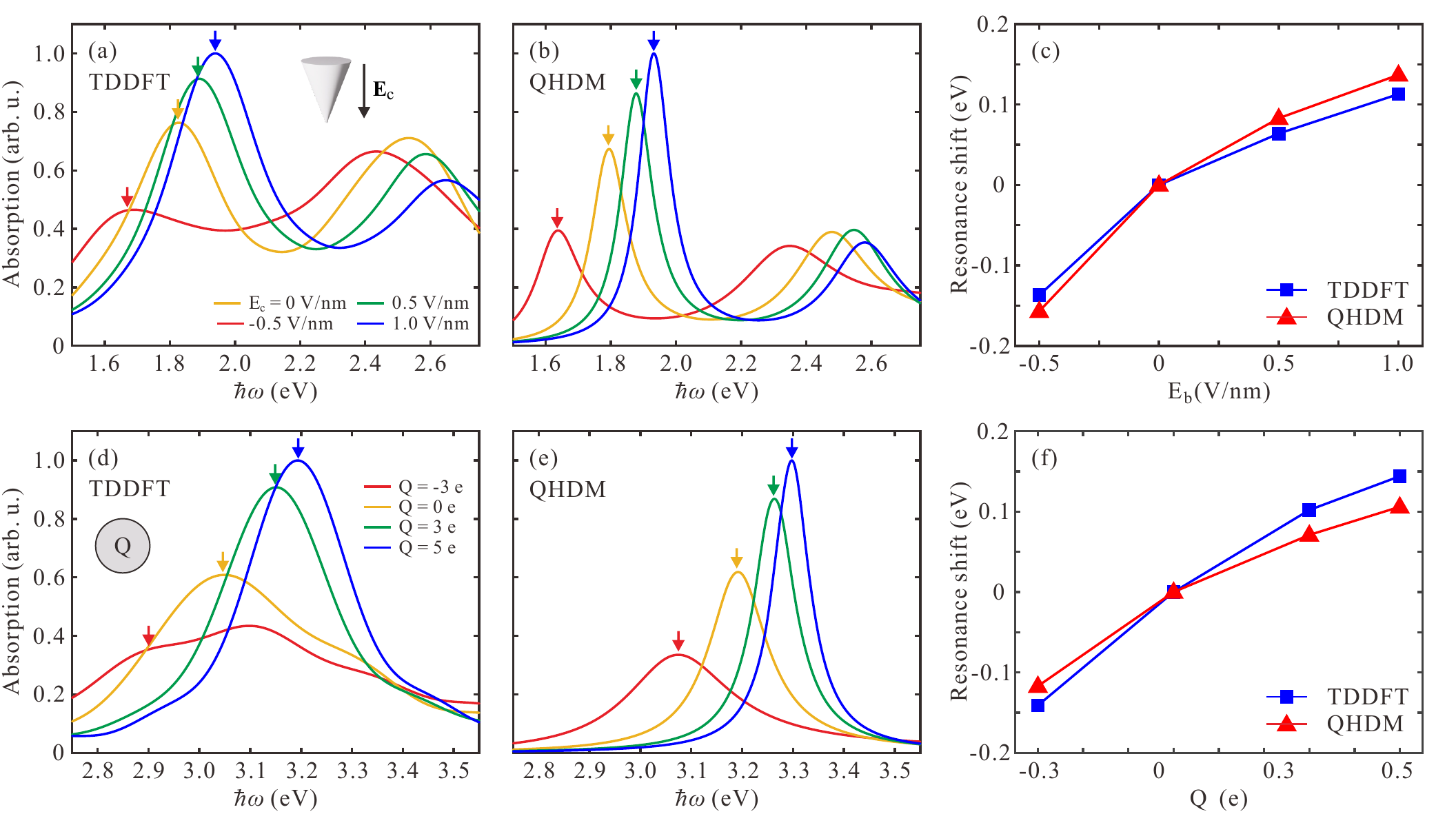}
\caption{Optical absorption spectra of
(a-c) a sodium nanocone biased with various electrostatic fields $\mathbf{E}_\mathrm{c}$ and (d-f) a sodium nanosphere charged with various net charges $Q$.
The spectra are calculated using (a, c) TD-DFT under jellium approximation and (b, e) the calibrated QHDM with $\lambda_\mathrm{w}=0.7$ and $\eta_\sigma=7$.
Frequency shifts of the main resonances (indicated by arrows) of the nanocone with respect to $\mathbf{E}_\mathrm{c}=0$ and nanosphere with respect to $Q=0$ are shown in (c) and (f) respectively.
\label{figV3SpecPlt2}}
\end{figure*}
\vspace*{.3cm}
\begin{figure*}[htb]
\centering
\includegraphics[width=0.95\textwidth]{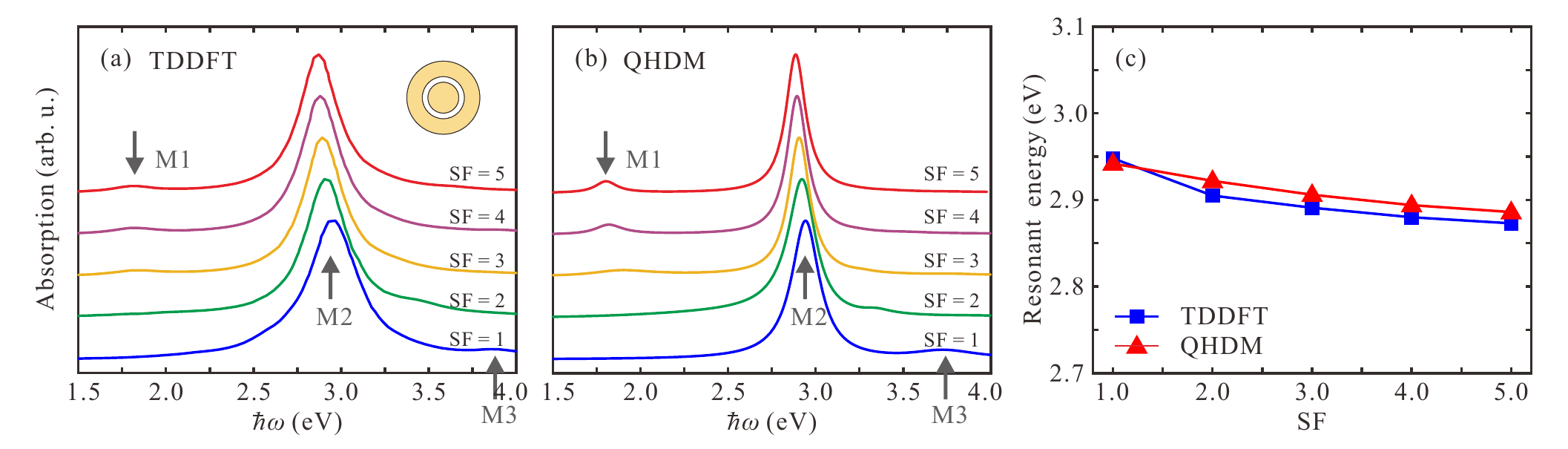}
\caption{
Application of the calibrated QHDM to the gold nanomatryoshka investigated in Fig.\,\ref{figV3StatPlt2}, where the significant effects of metal ion lattice are incorporated.
(a, b) Optical absorption spectra of the nanomatryoshkas of sizes 
$\mathrm{SF}\times(8.5,\allowbreak 9.5,\allowbreak 15.9)$ \AA\ 
calculated with TD-DFT (data adapted from Fig.\,(3) of Ref.\,\cite{,Kulkarni2013CoreShell}) and the calibrated QHDM respectively.
$\mathrm{SF}$ is a global geometrical scaling factor.
(c) The resonance energies of the major plasmon around $2.9$ eV.
M1, M2 and M3 denote the three dominant resonance modes.
\label{figV3SpecPlt3}}
\end{figure*}

\section{Calibration of linear-response QHDM}
Given the ground-state electron density distribution, we are in a position to calibrate the QHDM response calculation and to find the linear optical responses.
The QHDM response calculation involves two undetermined model parameters, i.e., the von Weizs\"acker parameter $\lambda_\mathrm{w}$ and strength of nonlocal damping $\eta_\sigma$.
Here $\lambda_\mathrm{w}$, different from that for ground state study, has influence on the frequencies of plasmon resonances predicted by QHDM.
On the other hand, $\eta_\sigma$ mainly influences the broadening and amplitudes of the resonances.
Following the same procedure for calibrating ground-state calculation, we fit these two parameters by studying the linear optical responses of sodium nanostructures.
Besides different geometries, we include in the calibration possible external electrostatic manipulation, e.g. electrical bias and charging, of the systems.
Practically the sizes of the nanostructures are limited by the computational load of TD-DFT.

We thus conceive two examples of a nanocone and a nanosphere subject to electrostatic manipulation.
Both the sodium nanocone with $45^\circ$ top angle and nanosphere are assumed to have in total $216$ conduction electrons.
An external control field $\mathbf{E}_\mathrm{c}$ is exerted along the axis to bias the nanocone.
For the sodium nanosphere, it is assumed electrical charged with the net charge $Q$.
The absorption spectra are systematically studied for the two nanoparticles under the influence of varying electrostatic control, i.e. $\mathbf{E}_\mathrm{c}$ and $Q$.
QHDM calculations are executed with $\lambda_\mathrm{w}$ and $\eta_\sigma$ running over the ranges shown in Fig.\,\ref{figV3SpecPlt1}.
The TD-DFT simulations are carried out using the open-source package Octopus \cite{,Rubio2020Octopus}, wherein the metal is treated under jellium approximation.
See appendix B for the details of the TD-DFT and QHDM response calculations.
In order to seek the best fitting of $\lambda_\mathrm{w}$ and $\eta_\sigma$, we define error functions to evaluate how faithfully the QHDM calculations agree with the TD-DFT results for the two systems.
The error functions have been designed to count the degrees of agreement in both the position and relative amplitude of the main plasmon resonances (see the colored arrows in Fig.\,\ref{figV3SpecPlt2}); see Appendix C for details.
The results of the systematic studies are then summarized in the plots of the error functions in Fig.\,\ref{figV3SpecPlt1}.
According to Fig.\,\ref{figV3SpecPlt1}(a), we have better predictions of QHDM with larger $\lambda_\mathrm{w}$, while the predictive power is insensitive to $\eta_\sigma$.
On the contrary, smaller $\lambda_\mathrm{w}$ is preferred according to Fig.\,\ref{figV3SpecPlt1}(b).
The value of $\eta_\sigma$ also becomes more critical.
Considering the error functions as a whole, the predictive power of QHDM response calculation is optimized with the choice of $(\lambda_\mathrm{w},\eta_\sigma)=(0.7, 7)$.
The pair is marked by the white stars in Fig.\,\ref{figV3SpecPlt1}.

As the confirmation of the parameter fitting, we explicitly illustrate in Fig.\,\ref{figV3SpecPlt2}(a,\,b) and (d,\,e) the complete absorption spectra for the two systems when the optimal model parameters are used.
In the two leftmost columns, we observe that the prominent features of the spectra by TD-DFT are successfully captured by QHDM calculations, including the positions, amplitudes and broadenings of the main resonances.
The absolute linewidths of the resonances by TD-DFT are apparently larger than those given by QHDM calculations.
That's because different phenomenological damping rates are adopted.
In the QHDM simulations, the damping rate is set as 
$\gamma\allowbreak =\allowbreak 0.066$ eV, 
whereas the rate is taken as 
\textcolor{black}{0.15} eV in TD-DFT calculations.
The agreement between QHDM and TD-DFT is especially satisfactory for the absorption spectra of the nanocone.
As shown in Fig.\,\ref{figV3SpecPlt2}(a,\,b), the evolution of an additional resonance around $2.4$ eV is also reproduced by QHDM at reasonably good accuracy.
Some subtleties arise in the resonances of the nanosphere by TD-DFT.
An unexpected resonance grows dominant when extra electrons are added.
The feature may be the result of the quantum effects associated with the electron orbitals, not available in QHDM.
Of particular interest is the resonance shifts caused by the electrostatic control.
In Fig.\,\ref{figV3SpecPlt2}(c,\,f), we evaluate and plot the shifts of the main plasmon resonances with respect to the resonance positions without electrostatic control.
Using the optimal model parameters, the QHDM and TD-DFT data largely agree with each other in the overall range of $\pm0.15$ eV.
The reliable QHDM response calculations for the above situations would facilitate novel applications of quantum plasmonics in optoelectronics \cite{,Marinica2015ActiveQuantumPlasmonics,Borisov2016Charge,Ludwig2020ActiveControl}.

\section{Refined QHDM responses of gold nanomatryoshkas}

\begin{figure}[b]
\centering
\includegraphics[width=0.48\textwidth]{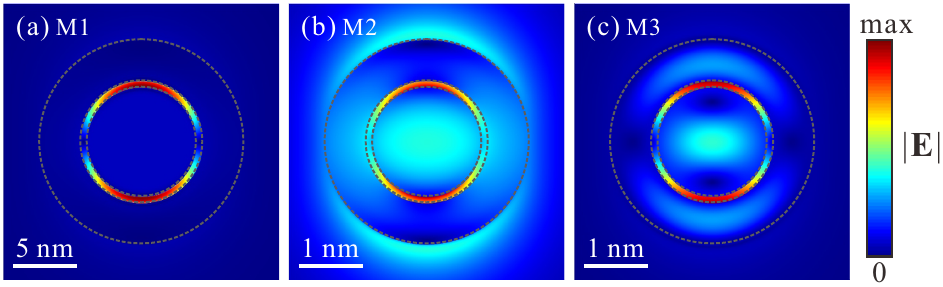}
\caption{
Electric field profiles $|\mathbf{E}|$ of the three dominant resonance modes in Fig.\,\ref{figV3SpecPlt3}(b).
(a) M1 mode of the nanomatryoshka with $\mathrm{SF}=5$.
(b, c) M2 and M3 modes of the nanomatryoshka with $\mathrm{SF}=1$.
\label{figV3SpecPlt4}}
\end{figure}

Although the key ingredients of the refined QHDM, or $\langle\delta V\rangle$ and $\varepsilon_{\mathrm{ml}}$ of metal, are only directly manifested in the ground-state formulation (see Section II), their influences on $n_0$ certainly would be reflected in the optical responses.
Here we demonstrate how the influences on $n_0$ are correctly relayed to the optical responses with the help of the above calibrated QHDM response calculation.
We have showcased the application of the calibrated ground-state QHDM to the gold nanomatryoshkas in Fig.\,\ref{figV3StatPlt2}.
The reported TD-DFT results for $n_0$ distribution have been faithfully repeated.
Then we continue the example and inspect their absorption spectra with the calibrated QHDM response calculation.

A series of absorption spectra are calculated for the gold nanomatryoshkas of sizes $\mathrm{SF}\!\times\!(8.5,9.5,15.9)$ \AA\ and compared with the TD-DFT data adapted from Fig.\,(3) of Ref.\,\cite{,Kulkarni2013CoreShell}.
$\mathrm{SF}\in\{1,2,3,4,5\}$ is a global geometrical scaling factor.
As shown in Fig.\,\ref{figV3SpecPlt3}(a, b), the absorption spectra predicted by QHDM show excellent agreement with the TD-DFT results, except that the resonance peaks in Fig.\,\ref{figV3SpecPlt3}(b) is sharper.
The reason is that the phenomenological damping rate $\gamma=0.135$ eV adopted in the QHDM simulations is smaller.
Our QHDM analysis successfully reveals the three resonance features around $1.8$, $2.9$ and $3.8$ eV.
For the major resonance around $2.9$ eV, we further examine its resonance frequency and depict it as a function of SF in Fig.\,\ref{figV3SpecPlt3}(c).
Continuous red shift is consistently reported by both methods.
Albeit the red shift is relatively weak ($\sim\!0.5\%$), the amount of the shift is semi-quantitatively predicted by the refined QHDM.
More remakably, the spectra calculated according to the refined QHDM exhibit the correct evolution trends of the two minor resonances.
We especially notice that the high-frequency resonance is absent in the spectra obtained with the conventional QHDM \cite{Khalid2018OpticalProperties}, and is captured only by using our refined theory.
Moreover, QHDM facilitates mode analysis \cite{Zhou2021General}, which is however not possible within TD-DFT.
We calculate the quasinormal modes dominating the resonances in Fig.\,\ref{figV3SpecPlt3} within the refined QHDM and display their electric field profiles in Fig.\,\ref{figV3SpecPlt4}.
The complex eigenfrequencies of the M1, M2 and M3 modes are $1.80+0.01i$ eV, $2.94+0.10i$ eV and $3.71+0.23i$ eV, respectively.
Among them, the M2 mode dominating the major resonance clearly exhibits much more optical field extending outside, leading to efficient access to far-field excitation and scattering.

\section{Conclusion}
In summary, we have carefully calibrated the refined QHDM developed by us 
\textcolor{black}{in a recent work \cite{,Li2021BiasModulation}.}
The involved model parameters, i.e., the von Weizs\"acker parameter $\lambda_\mathrm{w}$ and strength of nonlocal damping $\eta_\sigma$, have been determined by fitting TD-DFT results of sodium nanostructures.
Specifically, the $\lambda_\mathrm{w}=0.43$ ($\eta_\sigma$ is irrelevant) for ground-state calculation is extracted by examing the stationary electron density distribution and work function of a sodium slab.
For response calculation, $\lambda_\mathrm{w}=0.7$ and $\eta_\sigma=7$ are fixed through comprehensively evaluating the absorption spectra of a nanocone and a nanosphere under varying electrostatic control.
Furthermore we have deployed the calibrated QHDM to study the archetypical example of gold nanomatryoshkas.
Both the ground state and optical responses are thoroughly benchmarked against the reported TD-DFT data.
Excellent agreement has been observed in the $n_0$ distributions and absorption spectra.
Therefore the refined QHDM with the calibrated parameters proves to reliably predict stationary and optical properties of plasmonic nanostructures.
We thus envision it would serve as a valuable tool for nanoplasmonics researches.

\section*{Acknowledgement}
We acknowledge financial support from the National Natural Science Foundation of China (Grant Number 9215011 and 11874166) and Huazhong University of Science and Technology. The computing work in this paper is supported by the public computing service platform provided by the Network and Computing Center of HUST.

\section*{Appendices}
\subsection*{Appendix A: 1D DFT ground state calculation for metal slab}\label{AppendixA}
The DFT simulation follows the same prescription in Refs.\,\cite{,Fang2015QuantumSpillover,Neuhauser2012DFT1D} and our related work \cite{,Li2021BiasModulation}.
Specifically, the conduction electron density $n(x)$ is solved from the coupled KS equation and Poisson equation as follows,
\begin{gather}
\label{Eq:24:KSStat1D}
\left[-\frac{\hbar^2}{2m}\frac{d^2}{dx^2}
+ V_\mathrm{XC}(x) + q_\mathrm{e}\phi_0(x)
\right]\psi_j(x)
= \varepsilon_j\psi_j(x),
\\\label{Eq:25:Poisson1D}
\frac{d^2}{dx^2}\phi_0(x)
= \frac{q_\mathrm{e}}{\varepsilon_0}\left[n_+-n(x)\right].
\end{gather}
Here $\psi_j$ represents the $j$-th KS orbital of electron with the eigen-energy $\varepsilon_j$. 
The density $n(x)$ is constructed by summing $|\psi_j(x)|^2$ over the occupied orbitals up to the Fermi level $\mu$, which is determined by the charge neutrality condition.
The Wigner's XC potential $V_\mathrm{XC}$ \cite{,Liebsch1997ElectronicExcitations} and electrostatic potential $q_\mathrm{e}\phi_0$ above constitute the effective potential
$V_\mathrm{eff}=V_\mathrm{XC}+q_\mathrm{e}\phi_0$.
The work function then can be evaluated as 
$W=V_\mathrm{eff}(x\!\to\!\pm\infty)-\mu$.

\subsection*{Appendix B: TD-DFT and QHDM calculations of optical absorption spectrum}
We carry out the TD-DFT simulations using the open-source software Octopus \cite{,Rubio2020Octopus}, wherein the jellium approximation has been used to model metal nanoparticles.
The absorption spectra in Fig.\,\ref{figV3SpecPlt2} are obtained using the time-propagation approach and $\delta$-kick technique.
For the latter, the direction of the $\delta$-kick is set along the symmetry axes of the structures under consideration, i.e., the nanocone and nanosphere.
Concretely, we calculate the frequency-resolved absorption cross section spectrum $\sigma(\omega)$ by Fourier transforming the induced dipole moment with the Fourier transformation utility of Octopus.

In the QHDM simulations, the optical absorption from an incident electromagnetic field is obtained by solving Eqs.\,\eqref{Eq:15:HEqV3} and \eqref{Eq:16:WaveEq} in frequency domain, wherein the scattered-field formulation is invoked to rewrite the electric wave equation.
For a given frequency, the absorption power is evaluated by integrating the Poynting flux over a surface enclosing the scattering object, i.e., the nanocone or nanosphere in Fig.\,\ref{figV3SpecPlt3}.
The incidence is chosen to be a radially polarized beam \cite{,Li2021BrightOptical} propagating along the symmetry axis of the scattering object.

\subsection*{Appendix C: Error functions used for calibrating QHDM linear response}\label{AppendixB}
The error functions in Fig.\,\ref{figV3SpecPlt1} assess the faithfulness of the predictions by QHDM on the frequencies and relative amplitudes of the major plasmon resonances of the nanocone and nanosphere.
We shall respectively denote the frequencies and amplitudes corresponding to the parameter case $(\lambda_\mathrm{w},\eta_\sigma)$ of QHDM with
$\Omega_k^\alpha(\lambda_\mathrm{w},\eta_\sigma)$ and $F_k^\alpha(\lambda_\mathrm{w},\eta_\sigma)$,
where $\alpha$ indicates the structure (i.e., nanocone or nanosphere), and $k$ indicates the parameter of the electrostatic control.
For the nanocone, $k$ denotes the electrostatic control field $\mathbf{E}_\mathrm{c}$, with the field strength $E_\mathrm{c}\in\{-0.5,0,0.5,1.0\}$ V/nm.
For the nanosphere, $k$ denotes the net charge $Q$, with $Q\in\{-3,0,3,5\}\times e$.
The corresponding frequencies and amplitudes predicted by TD-DFT for given $\alpha$ and $k$ are respectively denoted with 
$\overline{\Omega}_k^\alpha$ and $\overline{\Omega}_k^\alpha$
, which are recognized as the reference.
All the amplitudes above are normalized by those in the cases without electrostatic control, i.e., $\mathbf{E}_\mathrm{c}=0$ for the nanocone, and $Q=0$ for the nanosphere.
Thereupon, we define for structure $\alpha$ the QHDM error function over the parameter cases $(\lambda_\mathrm{w},\eta_\sigma)$ as
\begin{align}
\mathrm{Err}^\alpha(\lambda_\mathrm{w},\eta_\sigma)
&= W_\Omega\,\sum_k\left[
\frac{\Omega_k^\alpha(\lambda_\mathrm{w},\eta_\sigma)-\overline{\Omega}_k^\alpha}{\overline{\Omega}_k^\alpha}
\right]^2
\notag\\
&+ W_F\,\sum_k\left[
\frac{F_k^\alpha(\lambda_\mathrm{w},\eta_\sigma)-\overline{F}_k^\alpha}{\overline{\Omega}_k^\alpha}
\right]^2,
\end{align}
where $W_\Omega$ and $W_F$ are two adjustable weighting factors.
We set 
$W_\Omega\allowbreak =\allowbreak W_F\times10$.
The error functions $\mathrm{Err}^\alpha(\lambda_\mathrm{w},\eta_\sigma)$ are finally normalized to ranges closed to $[0,1]$ (see Fig.\,\ref{figV3SpecPlt1}) with respect to some proper global constants.

\footnotetext[1]{Bias modulation paper.}
\bibliography{reference}

\end{document}